\documentclass[prd, aps, showpacs]{revtex4}

\usepackage{latexsym, graphicx}

\begin{document}

\title{Entanglement, recoherence and information flow in \\
an accelerated detector - quantum field system: \\
Implications for black hole information issue}
\author{Shih-Yuin Lin}
\email{sylin@phys.cts.nthu.edu.tw}
\affiliation{
Physics Division, National Center for Theoretical Sciences,
P.O. Box 2-131, Hsinchu 30013, Taiwan}
\author{B. L. Hu}
\email{blhu@umd.edu} \affiliation{Joint Quantum Institute and
Maryland Center for Fundamental Physics, Department of Physics,
University of Maryland, College Park, Maryland 20742-4111, USA
and \\
Perimeter Institute for Theoretical Physics,
31 Caroline Street North, Waterloo, Ontario N2L 2Y5 Canada}
\date{April 26, 2008}

\begin{abstract}
We study an exactly solvable model where an uniformly accelerated
detector is linearly coupled to a massless scalar field initially in
the Minkowski vacuum. Using the exact correlation functions obtained
in Refs. \cite{LH2005, LH2006} we show that as soon as the coupling
is switched on one can see information flowing from the detector to
the field and propagating with the radiation into null infinity. By
expressing the reduced density matrix of the detector in terms of the
two-point functions, we calculate the purity function in the detector
and study the evolution of quantum entanglement between the detector
and the field. Only in the ultraweak coupling regime could some
degree of recoherence in the detector appear at late times, but never
in full restoration. We explicitly show that under the most general
conditions the detector never recovers its quantum coherence  and the
entanglement between the detector and the field remains large at late
times. To the extent this model can be used as an analog to the
system of a black hole interacting with a quantum field, 
our result seems to suggest in the prevalent non-Markovian regime,
assuming unitarity for the combined system, that  black hole
information is not lost but transferred to the quantum field degrees
of freedom. Our combined system will evolve into a highly entangled
state between a remnant of large area (in Bekenstein's black hole
atom analog) without any information of its initial state, and the
quantum field, now imbued with complex information content
not-so-easily retrievable by a local observer.
\end{abstract}

\pacs{04.62.+v, 
42.50.Lc 
04.70.Dy} 

\maketitle


\section{Introduction}

Recent years saw increased interest and activities in understanding
the nature and dynamics of entanglement in quantum systems. This is a
distinctly quantum feature absent in classical systems, also the most
essential factor in demonstrating the capability of any proposed
quantum computer scheme.  Many theoretical studies of quantum
entanglement have been carried out for two level systems in atoms,
quantum dots and superconductors with charges, spins and fluxes.
Entanglement between qubits (quantum two-level systems) as mediated
by a quantum fields is also of interest because of realistic
experimental needs. This aspect has been considered by one of us and
collaborators in popular models such as atoms in a cavity field (see
\cite{YuEbe04, YuEbe02, YuEbe03} for two qubits in two separate 
fields and \cite{RRS05, RCR05, FicTan06, FicTan02, ASH07, vSM07}         
in one quantum field) and 
two Brownian oscillators in a finite temperature bath (see, e.g.,
\cite{CYH07a,CHY07b} and references therein). Associated with the
dynamics of entanglement are the older issues of quantum decoherence
and entropy / information evolution. Entanglement between two qubits
is measured effectively by the concurrence functions \cite{Wootters},
the entropy of formation, fidelity and purity  (see, e.g.,
\cite{Viola04, Viola03}). Investigation of these issues are pursued 
most actively today in the area of quantum information.

Another area where questions of quantum entanglement, entropy and
information are raised is in black hole physics and quantum gravity,
where fundamental issues about the laws of physics and the structure
of spacetime are of interest. Bekenstein's proposal of black hole
entropy \cite{Bek73,Bek94} and Hawking's discovery
\cite{Haw74,Haw75,Haw76a} that black hole emits thermal radiance
raise the question of how the end state of a black hole is like, and
ushered in the debate whether there is net information lost in a
black hole  \cite{Haw76b,Haw05,Page80} (see, e.g.,
\cite{Preskill93,Page94} for an overview) and, by proxy, whether
unitarity in the laws of physics is violated. There are many proposed
explanations -- remnants,  naked singularity, baby universe formation
or complete evaporation. (See e.g., \cite{Witten91,CGHS,RST92a,RST92b,
HolWil92, Wilczek93,PolStr}.) We will focus on 
one explanation which bears on some of the issues in quantum coherence
and information mentioned above, namely, that information in the
black hole is not lost but transferred and dispersed into the quantum
field through Hawking radiation. Many authors have proposed or hinted
on this view, approached from different angles, some more explicitly
than others, ranging from quantum gravity, string theory,
nonequilibrium statistical mechanics and quantum information. Out of
these we shall focus on the themes of two papers
\cite{AngRecoh, HuErice} (which are shared in varying degrees by other
authors, see papers cited above). For some recent discussions with
newer perspectives, see e.g., \cite{LowTho,Gid06} from string theory
and nonlocality viewpoints, \cite{Ter05} from loop quantum gravity and 
\cite{SmoOpp,BraPat} for a quantum information approach to this issue.

In Ref. \cite{AngRecoh} Anglin {\it et al.} studied a simple model of a
harmonic oscillator coupled to a quantum field and analyzed the
decoherence and recoherence of the detector in the ultraweak coupling
approximation. They claimed that, as soon as the coupling is switched
on, the oscillator loses the quantum coherence on a very short
decoherence time corresponding to the cut-off time scale. But all the
quantum coherence will recover in the end, after a much longer
relaxation time. The initial information in the oscillator eventually
flow into the field side, though appearing in a highly nontrivial
form. They suggest that black hole information could behave in a
similar way. In Ref. \cite{HuErice} one of us described a scenario
where information in the black hole is transferred to the quantum
field, and proposed to use the correlation functions of an
interacting field as registers of information and the dynamics of
correlations as a measure of information flow. According to the
viewpoint put forth in \cite{HuErice}, the appearance of information
loss is primarily owing to the fact that actual physical measurements
are limited in accuracy, i.e., one can only access the lowest order
correlation functions, beginning with the mean field and the
two-point functions. It also highlights the huge capacity of a
quantum field in storing and dispensing information.

In our present work we adopt some modern methods of quantum
information to address some aspects of black hole information issues.
Looking at a particle detector in uniform acceleration interacting
with a quantum field, we study the quantum entanglement, quantum
recoherence and information flow in this simple system with the aim
of providing an illustrative example to check on the ideas put forth
in \cite{HuErice}. Of the main points made there, we can confirm that
information does flow from the detector to the field and gets
propagated to null infinity with the radiation. Further, from the
behavior of two-point functions obtained in \cite{LH2005, LH2006}, we
do not see any qualitative change between the $a=0$ static case and
the $a \neq 0$ uniformly accelerated cases. Insofar as the
information behavior is concerned the uniformly accelerated cases are
closer in nature to the inertial case than those cases with
non-uniform acceleration \cite{Koks, HuPJCapri}.
We also analyze the cases similar to those considered in
\cite{AngRecoh} but with a much wider parameter range. We clarify
that, in the ultraweak coupling limit, some level of quantum
coherence is recovered at late times, but full recoherence is
impossible as long as the coupling is on; Beyond the ultraweak
coupling limit, late-time recoherence never occurs.
In this study we show that with new tools adopted from quantum
information, even simple models like this can provide valuable
insight into deeper issues in statistical mechanical properties of
detector (atom)- field interactions and by suitable extension, black
hole spacetime physics.

This paper is organized as follows: In Sec. II we introduce a system
with the uniformly accelerated detector (atom) coupled with the field.
Using the exact form of the two-point functions we demonstrate that
already at this stage one can see information flowing from the
detector to the field and propagates with the radiation into null
infinity. In Sec IIA we introduce a quantity called purity and show
that it is useful for describing the entanglement and the recoherence
dynamics of the system. To obtain the purity function we calculate
the reduced density matrix of the detector in Sec. IIB. Then in
Sec.III we derive analytic expressions for the purity in the detector
and deduce the entanglement between the detector and the field, first
from the detector's view (time) and then from the field's view
(time). This behavior is illustrated with graphs plotted in various
parameter ranges. In Sec. IV we make some technical remarks and then
give a discussion of the central issues raised here. As a
complementary description, in Appendix A, we give the reduced density
matrix of the field for the same initial state used in Sec. II. For a
detector initially in the ground state the purity of the detector and
the purity of the field have simple enough expressions that we can
record them in Appendix B. Finally in Appendix C we provide
a comparison of our results to the claims of \cite{AngRecoh} on the
recoherence issue.

\section{Information flow between the detector and the quantum field}
\label{EvsC}

Consider a harmonic oscillator with bare mass $m_0$,  bare natural
frequency $\Omega_0$ and internal degree of freedom $Q$ (such as the
Unruh-DeWitt(UD) detector \cite{Unr76, DeW79, BD} in the language of
quantum field theory in curved spacetime) interacting with a massless
quantum scalar field $\Phi$ in four-dimensional Minkowski space with
coupling constant $\lambda_0$. The action of the combined particle
detector - quantum field system is given by \cite{LH2005}
\begin{eqnarray}
  S &=&\int d\tau {m_0\over 2}\left[ \left(\partial_\tau Q\right)^2
    -\Omega_0^2 Q^2\right] -\int d^4 x {1\over 2}\partial_\mu\Phi
    \partial^\mu\Phi\nonumber\\  
    & &+{\lambda_0}\int d\tau\int d^4 x Q(\tau)\Phi (x)
  \delta^4\left(x^{\mu}-z^{\mu}(\tau)\right). \label{Stot1}
\end{eqnarray}
We will consider the cases when it is uniformly accelerated along the
trajectory $z^\mu(\tau)=(a^{-1}\sinh a\tau, a^{-1}\cosh a\tau,0,0)$
with proper acceleration $a$. For the cases of detectors at rest
($a=0$), we have learnt in \cite{LH2005, LH2006} that the two-point
functions of our UD detector theory in (3+1)D with finite $a$ have no
singular behavior as $a\to 0$. Hence all our results below, which are
expressed in terms of these two-point functions, apply equally well
to the case of detectors at rest.

We study the case when the initial state of the combined system is a
direct product of a quantum state $\left|\right. q\left.\right>$ for
the detector $Q$ and the Minkowski vacuum $\left|\right. 0_M
\left.\right>$ for the field $\Phi$,
\begin{equation}
  \left|\right. \psi(\tau_0)\left.\right> =
  \left|\right. q\left.\right> \otimes
  \left|\right. 0_M \left.\right>. \label{initstat}
\end{equation}
Oftentimes one hears the comment that since the detector is a
harmonic oscillator and the field can be decomposed as a collection
of harmonic oscillators, the combined system is just a system of
$(\infty+1)$ oscillators.  Indeed one can diagonalize the UD
detector-field system into a system of $(\infty+1)$ free harmonic
oscillators \cite{AngRecoh}. But we want to add the warning that the
choice of initial conditions for the detector $Q$  (such as the
ground state) and for the field $\Phi$ (such as the Minkowski vacuum)
separately renders the set-up of this system (physical variables and
initial states) physically different from that of the system of
$(\infty+1)$ free harmonic oscillators with choices of the initial
states deemed natural for it.

Since the combined system is linear, the operators evolve in the
Heisenberg picture as linear transformations. For example, the
operator of $Q$ evolves like \cite{LH2005}
\begin{eqnarray}
  \hat{Q}(\tau) &=& \sqrt{\hbar\over 2\Omega_r m_0}\left[
    q^a(\tau)\hat{a}+q^{a*}(\tau)\hat{a}^\dagger \right] \nonumber\\
    & & + \int {d^3 k\over (2\pi)^3}\sqrt{\hbar\over 2\omega}
    \left[q^{(+)}(\tau,{\bf k})\hat{b}_{\bf k} +
    q^{(-)}(\tau,{\bf k})\hat{b}_{\bf k}^\dagger\right],
\end{eqnarray}
where $q^a$ and $q^{(\pm)}$ are c-number functions, $\hat{a}$,
$\hat{a}^\dagger$ are the lowering and the raising operators of the
free detector, and $\hat{b}_{\bf k}$, $\hat{b}_{\bf k}^\dagger$ are  
the annihilation and creation operators of the free field. When 
sandwiched by the factorized initial state $(\ref{initstat})$, the 
two-point functions of the detector and those of the field split 
into two parts, e.g.,
\begin{equation}
  \left<\right. Q(\tau)Q(\tau')\left.\right>  =
  \left<\right.q \,|\, q\left.\right>\left< \right. Q(\tau)Q(\tau')
  \left.\right>_{\rm v}+ \left< \right.Q(\tau)Q(\tau')\left.
  \right>_{\rm a}\left< 0_M| 0_M\right>. \label{splitQQ}
\end{equation}
with
\begin{eqnarray}
  & &\left< \right. Q(\tau)Q(\tau')\left.\right>_{\rm v} =\nonumber\\
  & & \,\,\, \left< 0_M\right.|\int {d^3 k\over (2\pi)^3}
    \sqrt{\hbar\over 2\omega}q^{(+)}(\tau,{\bf k})\hat{b}_{\bf k}
    \int {d^3 k'\over (2\pi)^3}\sqrt{\hbar\over 2\omega'}
    q^{(-)}(\tau',{\bf k'})\hat{b}_{\bf k'}^\dagger|
    \left. 0_M\right>, \label{defQQv}\\
  & &\left< \right.Q(\tau)Q(\tau') \left.\right>_{\rm a} =\nonumber\\ 
    & & \,\,\, {\hbar\over 2\Omega_r m_0}
    \left< \right. q\left.\right| \left[
    q^a(\tau)\hat{a}+q^{a*}(\tau)\hat{a}^\dagger \right]
   \left[ q^a(\tau')\hat{a}+q^{a*}(\tau')\hat{a}^\dagger \right]
    \left|\right.q\left. \right>. \label{defQQa}
\end{eqnarray}
Here $\left<\right. ..\left.\right>_{\rm a}$ depends on the initial
state of the detector only, while $\left<\right. ..\left.
\right>_{\rm v}$ depends on the initial state of the field, namely
the Minkowski vacuum. Therefore by studying the correlation functions
$\left<\right. .. \left.\right>_{\rm a}$ of the detector and those of
the field, one can monitor how the information initially in the
detector is flowing into the field.

Indeed, from Ref. \cite{LH2005}, we learned that $\left<\right.
Q^2\left. \right>_{\rm a}$, $\left<\right.P^2\left.\right>_{\rm a}$
and $\left<\right.P,Q\left.\right>_{\rm a}$ all decay after the
coupling is switched on, and the information about the initial state
of the detector is subsumed into the quantum field (in
$\left<\right.\Phi(x)\Phi(x')\left. \right>_{\rm a}$, etc.) This view
is further supported by the the energy conservation law \cite{LH2005}
\begin{equation}
  \dot{E}(\tau) = -2\gamma m_0 \left<\right.\dot{Q}^2(\tau)
  \left.\right>_{\rm tot}. \label{Econserv}
\end{equation}
between the internal energy of the detector (left hand side) and
the radiated energy of a monopole radiation (right hand side). Here
$\gamma\equiv \lambda_0^2/8\pi m_0$ and
\begin{equation}
  E(\tau) \equiv \left< {1\over 2m_0} P^2+
          {1\over 2}m_0 \Omega_r^2 Q^2 \right>
\end{equation}
is the expectation value of the Hamiltonian of the detector,
$\left<\right.\dot{Q}^2(\tau)\left.\right>_{\rm tot}$ is the sum of
$\left<\right.\dot{Q}^2(\tau)\left.\right>$ and the interfering term
involving vacuum fluctuations of the field \cite{LH2005}.
From $(\ref{splitQQ})$ and $(\ref{Econserv})$, one sees that the
energy in $\left<\right.Q^2\left.\right>_{\rm a}$ and
$\left<\right.P^2\left.\right>_{\rm a}$ will be converted to monopole
radiation while the state of the detector at late times is sustained                 
only by the vacuum fluctuations of the field, namely, the reduced      
density matrix will be described by $\left<\right. .. \left.
\right>_{\rm v}$ only.

\subsection{Purity as a measure of quantum coherence and entanglement}
\label{puritypure}

For the consideration of quantum coherence and entanglement, one
needs to go beyond single one-point or two-point functions. The
quantity which contains all the information relevant to what we
want to investigate here is the reduced density matrix (RDM)
$\rho^R$, that of the detector when the field degrees of freedom
are integrated over, and the RDM of the field when that of the
detector is integrated over. Quantum information will be
extracted from $Tr (\rho^R)^n$, $n>1$, which can be expressed as
combinations of correlation functions.

Suppose a bipartite system, such as our detector-field system, is
in a pure state described by the normalized wave function(al)
$\psi [\Phi,Q]$. The density matrix for the combined system in
$(\Phi,Q)$ representation reads
\begin{equation}
  \rho [\Phi,Q;\Phi',Q'] = \psi[\Phi,Q]\psi^*[\Phi',Q'].
\end{equation}
It is always possible to perform a Schmidt decomposition on the
quantum state for the combined system so that \cite{PeresQT}
\begin{equation}
  \psi [\Phi,Q] = \sum_{j=1}^N M_j \Psi_j[\Phi] \varphi_j (Q),
\end{equation}
where $\Psi_j[\Phi]$ and $\varphi_j (Q)$ are orthonormal states of
the field and the detector, respectively. $M_j$'s are non-vanishing
coefficients, $N$ in total, chosen to be real here. $N$ is at 
most equal to the smaller of the dimensionalities of the two subsystems.
If $N>1$, $\psi[\Phi,Q]$ is an entangled (non-separable) 
state.

The RDM of the detector
\begin{equation}
  \rho^R(Q,Q') = Tr_\Phi \, \rho =
    \sum_{j=1}^N M_j^2 \varphi_j (Q) \varphi_j^* (Q'),
\label{schmidt}
\end{equation}
and the RDM of the field
\begin{equation}
  \rho^R[\Phi,\Phi'] = Tr_Q \, \rho =
    \sum_{j=1}^N M_j^2 \Psi_j [\Phi] \Psi_j^* [\Phi'],          
\end{equation}
share the same eigenvalues $M_j^2$.
Therefore
\begin{equation}
  Tr_Q \left[\rho^R(Q,Q')\right]^n = Tr_\Phi \left(\rho^R
  [\Phi,\Phi']\right)^n = \sum_{j=1}^N M_j^{2n} \label{PPequal}
\end{equation}
have the same values from both RDMs for all $n=1,2,3,\ldots$. In
particular, the purity ($n=2$) of the detector has the same value as
the purity of the field. The von Neumann entropy $\sim -\rho^R \log
\rho^R = -\sum_{j=1}^N M_j^2 \log M_j^2$ of the detector and of the
field are also the same. Note that the {\it form} of the equality    
$(\ref{PPequal})$ holds for every choice of time slice where the 
Hamiltonian for the combined system is defined. The {\it values} of 
$Tr (\rho^R)^n$ for different choice of time slice can be different.      

Clearly the purity of each sub-system measures the entanglement
between them: For $N=1$, $\psi$ is factorizable and ${\cal P}\equiv
Tr_Q \left[\rho^R(Q,Q')\right]^2 = {\cal P}_\Phi \equiv
Tr_\Phi\left(\rho^R [\Phi,\Phi']\right)^2=1$, while for $N>1$, $\psi$
is entangled and ${\cal P}={\cal P}_\Phi <1$. On the other hand, the
purity of a two-level atom is proportional to its polarization, thus
providing a measure of quantum coherence in that atom. 
Indeed, a two-level atom is not fully decohered until it becomes an       
infinite-temperature thermal state where the purity is in its least         %
value $1/2$. A thermal state at finite temperature, though looks classical, %
is not fully decohered and the purity is always greater than $1/2$.         %
(Note that a diagonal RDM can nonetheless possess nonzero off-diagonal      %
elements under a basis transformation.) Here we extend this view to our     %
system and use the value of the purity function as a measure of quantum   
coherence in the detector and in the field.
(Note that ``quantum coherence" used here does not refer to the 
off-diagonal elements of the density matrix, known to some authors 
as the ``coherences" \cite{API}.)

The behavior of quantum coherence ``flow" is quite different from energy 
flow. When the coupling is switched on, both the quantum coherence in 
the detector and the quantum coherence in the field decrease, while the
entanglement between them increases. So quantum coherence does not
flow from one subsystem to the other; It goes into sustaining the 
entanglement between the two subsystems \cite{Ter05}.                   

\subsection{RDM of the detector initially in the cat state}
\label{giveRDMD}

As an example, let us consider the detector in a cat state at the
initial moment $\tau_0$,
\begin{equation}
  \left|\right. q(\tau_0)\left.\right> =
  \cos \varphi\left|\right. E_0\left.\right>+
  e^{i\delta}\sin\varphi\left|\right. E_1\left.\right>,
  \label{initcat}
\end{equation}
where $\left|\right. E_0\left. \right>$ and $\left|\right.
E_1\left.\right>$ are the ground state and the first excited state of
the free detector, $\varphi$ is the mixing angle and $\delta$ is a
constant phase. The RDM of the detector for the initial state
$(\ref{initcat})$ reads
\begin{equation}
  \rho^R (Q,Q';\tau)
  = \rho^R_{(00)}\cos^2\varphi +\rho^R_{(11)}\sin^2\varphi +
  \left(e^{i\delta}\rho^R_{(10)}+e^{-i\delta}\rho^R_{(01)}\right)
  \sin\varphi\cos\varphi   \label{RDMQcat}
\end{equation}
where
\begin{equation}
  \rho^R_{(mn)} (Q,Q';\tau)\equiv \int {\cal D}\Phi_{\rm k}
  \psi_m [Q,\Phi_{\rm k};\tau] \psi_n^* [Q',\Phi_{\rm k};\tau],
\end{equation}
with $m,n=0,1$. Here $\psi_0$ is the wave functional corresponding to
an initial state given by the tensor product of the Minkowski vacuum
and the ground state, and $\psi_1$, the product with the first
excited state of the detector. In Schr\"odinger representation, one
has
\begin{eqnarray}
  & &\rho^R(Q,Q';\tau) = \sqrt{G^{11}+G^{22}+2G^{12}\over\pi}
    e^{-G^{ij}Q_i Q_j}\times \nonumber\\ & & \,\,\, \left\{\cos^2\varphi 
    +\sin^2\varphi(C +A^{ij}Q_i Q_j ) + \right.\nonumber\\ & & 
    \,\,\,\,\,\, \left.\sin\varphi\cos\varphi
    \left[(e^{i\delta}B^1+e^{-i\delta}B^{2*})Q+
  (e^{-i\delta}B^{1*}+e^{i\delta}B^2) Q'\right]\right\}, \label{RDMD}
\end{eqnarray}
where $i,j =1,2$, $Q_i = (Q, Q')$. The coefficients $C$, $A^{ij}$, 
$B^j$, and $G^{ij}$ could be expressed in terms of the correlation 
functions of the detector by looking at
\begin{eqnarray}
  \left<\right.Q^2 \left.\right>_{mn} &=& Tr[ Q^2 \rho^R_{(mn)} ], \\
  \left<\right. P,Q \left.\right>_{mn} &\equiv& {1\over 2}
    \left<\right. (PQ + QP)\left.\right>_{mn} =  {1\over 2i}
    Tr[ Q(\partial_{Q}-\partial_{Q'}) \rho^R_{(mn)} ], \\
  \left<\right. P^2 \left.\right>_{mn} &=& -{1\over 4} Tr[
  (\partial_Q^2+\partial_{Q'}^2 -2 \partial_Q\partial_{Q'})\rho^R_{(mn)}],
\end{eqnarray}
and so on, where $\left<\right. ..\left.\right>_{mn} \equiv
\left<\right. E_m, 0_M |\, .. \, |E_n, 0_M \left.\right>$. Explicit
expressions of the two-point functions
needed in this paper have been listed in Appendix A of Ref. \cite{LH2006}.

The elements of $G^{ij}$ are still the same as those in \cite{LH2005,
LH2006} for the detector initially in the ground state. Expressed in
terms of the two-point functions, they are given by
\begin{eqnarray}
  & & G^{11}+G^{22}+2G^{12} = {1\over 2 \left<\right. Q^2
    \left.\right>_{00}}, \label{G1}\\
  & & G^{11}+G^{22}-2G^{12} = {2\over \hbar^2\left<\right. Q^2
    \left.\right>_{00}} \left[\left<\right. P^2 \left.\right>_{00}
    \left<\right. Q^2 \left.\right>_{00} - \left(\left<\right. P,Q
    \left.\right>_{00}\right)^2 \right] ,\label{G2}\\
  & & G^{11}- G^{22} = -{i \left<\right. P,Q \left.\right>_{00}
    \over \hbar \left<\right. Q^2 \left.\right>_{00}}. \label{G3}
\end{eqnarray}
In writing $(\ref{RDMD})$ we have chosen a normalization such that
\begin{equation}
  C = \left<\right. Q^2\left.\right>_{\rm v}/\left<\right. Q^2
    \left.\right>_{00},
\end{equation}
which is zero at the initial moment $\tau=\tau_0$ and unity at late
times. From $(\ref{splitQQ})$, $(\ref{defQQv})$ and $(\ref{defQQa})$,
we have $\left<\right. .. \left.\right>_{00} = \left<\right. ..
\left.\right>_{\rm a_0} + \left<\right. .. \left.\right>_{\rm v}$ and
$\left<\right. .. \left.\right>_{11} = 3\left<\right. ..
\left.\right>_{\rm a_0} + \left<\right. .. \left.\right>_{\rm v}$,
where $\left<\right. .. \left.\right>_{\rm a_0} \equiv
\left<\right.E_0\, | .. |\,E_0 \left.\right>$ are the same as those
$\left<\right. .. \left. \right>_{\rm a}$ in \cite{LH2005,LH2006}.
Then the coefficients $A^{ij}$ in $(\ref{RDMD})$ are obtained by
solving
\begin{eqnarray}
  & & A^{11}+A^{22}+2A^{12} =
    {\left<\right. Q^2\left.\right>_{\rm a_0} \over \left(
    \left<\right. Q^2\left.\right>_{00}\right)^2}, \\
  & & A^{11}+A^{22}-2A^{12} = {4 \over \hbar^2 } \left \{
    -\left<\right. P^2\left.\right>_{\rm a_0} + \right. \nonumber\\ 
    & & \,\,\,\left. {\left<\right. P,Q\left.\right>_{00}\over
    \left(\left<\right. Q^2\left.\right>_{00}\right)^2}\left[
    2\left<\right.P,Q\left.\right>_{\rm a_0}\left<\right. Q^2
    \left.\right>_{00} -\left<\right. Q^2\left.\right>_{\rm a_0}
    \left<\right.P,Q\left.\right>_{00}\right]\right\},\\
  & & A^{11} -A^{22}=
  -{2i\over\hbar\left(\left<\right. Q^2\left.\right>_{00}\right)^2}
    \left[\left<\right. P,Q\left.\right>_{\rm v}\left<\right. Q^2
    \left.\right>_{\rm a_0}-\left<\right. P,Q\left.\right>_{\rm a_0}
    \left<\right. Q^2\left.\right>_{\rm v}\right].
\end{eqnarray}
It is known that $\left<\right. .. \left.\right>_{\rm a_0} \propto
e^{-2\gamma\eta}$ where $\eta\equiv \tau-\tau_0$ \cite{LH2005,LH2006}, 
so all $A^{ij}$ vanish at late times.

The coefficients $B^i$ can be easily found in a similar way:
\begin{eqnarray}
  B^1+B^2 &=& {\left<\right. Q\left.\right>_{10}
    \over \left<\right. Q^2 \left.\right>_{00}},\\
  B^1-B^2 &=& {2i\over\hbar}
    \left(\left<\right. P\left.\right>_{10}-
    {\left<\right. P,Q \left.\right>_{00}\over \left<\right. Q^2
    \left.\right>_{00}}\left<\right. Q\left.\right>_{10}\right).
\end{eqnarray}
In the Heisenberg picture, one has
\begin{eqnarray}
  \left<\right. Q\left.\right>_{10} &\equiv& \left<\right. 0_M |_\Phi
    \left<\right. E_1 |_Q \hat{Q}(\tau) | E_0\left.\right>_Q
    |0_M \left.\right>_\Phi = \sqrt{\hbar\over 2\Omega_r m_0}
      q^{\rm a*}(\tau),\\ \left<\right. P\left.\right>_{10}
    &=& m_0 \left<\right. \dot{Q} \left.\right>_{10} =
    \sqrt{\hbar m_0\over 2\Omega_r}\dot{q}^{\rm a*}(\tau),
\end{eqnarray}
where, as found in Ref. \cite{LH2005},
\begin{equation}
  q^a(\eta) = {1\over 2}\theta(\eta)e^{-\gamma\eta}
  \left[\left(1-{\Omega_r+ i\gamma\over\Omega}\right)e^{i\Omega\eta}+
  \left(1+{\Omega_r+i\gamma\over\Omega}\right)e^{-i\Omega\eta}\right],
\label{qa}
\end{equation}
where $\Omega_r$ is the renormalized natural frequency of the detector
and $\Omega\equiv \sqrt{\Omega_r^2 -\gamma^2}$.
Thus both $\left<\right. Q\left.\right>_{10}$ and $\left<\right.
P\left.\right>_{10}$ are proportional to $e^{-\gamma\eta}$ and
both $\rho^R_{(10)}$ and $\rho^R_{(01)}$ vanish at late times, when
\begin{equation}
  \left.\rho^R\right|_{\gamma\eta \gg 1}= \left.\rho^R_{(00)}
  \right|_{\gamma\eta \gg 1}\label{lateRhoR}
\end{equation}
for all choices of $\delta$ and $\varphi$ for initial states.

\section{Entanglement between the detector and the field}

\subsection{View from the detector}

From $(\ref{RDMD})$, the purity of the detector for the initial state
$(\ref{initcat})$ is given by
\begin{eqnarray}
& &{\cal P}(\tau) = {\hbar\over 2{\cal U}} \left( \cos^2\varphi +
{\left<\right.Q^2 \left.\right>_{\rm v} \over \left<\right. Q^2 \left.
\right>_{00}}\sin^2\varphi\right)^2 \nonumber\\
&+&{\hbar^3 \left<\right. Q^2 \left.\right>_{00}^2\over 4{\cal U}^3}
\sin^2\varphi\left[\left( \cos^2\varphi +  {\left<\right. Q^2
\left.\right>_{\rm v} \over \left<\right. Q^2 \left.\right>_{00}}
\sin^2\varphi\right) 
\left( 2{\rm Tr}A \,{\rm Tr}G-8 A^{12}G^{12}\right)\right.\nonumber\\
& & \left. + 2\cos^2\varphi \left( \left|B_\delta\right|^2{\rm Tr}A-
2G^{12}{\rm Re}\left(B_\delta\right)^2\right)\right]\nonumber\\
&+&{\hbar^5 \left<\right. Q^2 \left.\right>_{00}^4\over 8{\cal U}^5}
\sin^4\varphi\left[ 6A^{11}A^{22}\left({\rm Tr}G\right)^2
-24 A^{12}G^{12}{\rm Tr}A \, {\rm Tr}G \right. \nonumber\\ & &  
+\left.\left(\left(A^{11}\right)^2+ \left(A^{22}\right)^2+
4\left(A^{12}\right)^2\right)\left(\left({\rm Tr}G\right)^2+
8 \left(G^{12}\right)^2\right)\right],\label{purecat}
\end{eqnarray}
where ${\rm Tr}A = A^{11}+A^{22}$, ${\rm Tr}G = G^{11}+G^{22}$, 
$B_\delta\equiv e^{i\delta}B^1+e^{-i\delta}B^{2*}$ and ${\cal U}$ is 
defined in $(\ref{UncertFn})$.
As shown in Section \ref{puritypure} the purity of the field at the
moment $t=a^{-1}\sinh a\tau$ has the same value as $(\ref{purecat})$.
At late times ($\eta \equiv \tau-\tau_0\gg 1/\gamma$), as the
coefficients $A^{ij}$ and $B^j$ die out and the RDM is becoming
$(\ref{lateRhoR})$, the purity goes to [See $(\ref{purity00})$],
\begin{eqnarray}
  & & {\cal P}|_{\gamma\eta\gg 1} =
  \left.{\hbar\over 2{\cal U}}\right|_{\gamma\eta \gg 1}
    \nonumber\\ & & = \pi\Omega\left\{ {\rm Re}\,
    \left[{ia\over \gamma+i\Omega}-2i\psi_{\gamma+i\Omega}\right]
    \right\}^{-1/2}\times\nonumber\\ & & \,\,\,\,\, \left\{\left(
  {\rm Re}\,\left\{ (\Omega-i\gamma)^2 \left[ {ia\over \gamma+i\Omega}-
  2i\psi_{\gamma+i\Omega}\right]\right\} + 4 \gamma \Omega
  \left[\Lambda_1  -\ln {a\over\Omega}\right]\right)\right\}^{-1/2},
\label{lateP}
\end{eqnarray}
whose value is determined by the vacuum fluctuations, for {\it all}
initial states of the detector \cite{LH2005}. Here $\psi_s \equiv
\psi(1 + a^{-1}s)$ is the digamma function.

\begin{figure}
\includegraphics[width=5cm]{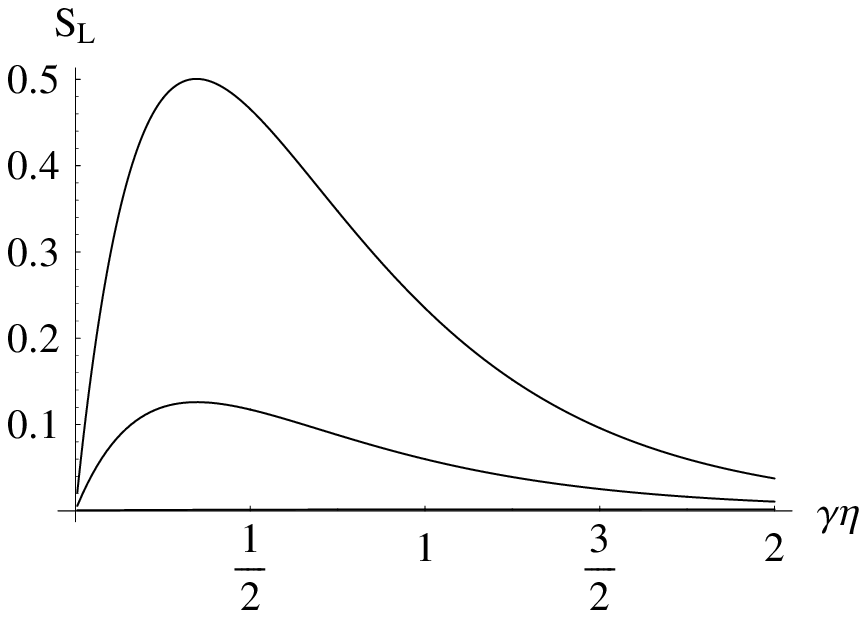}
\includegraphics[width=5cm]{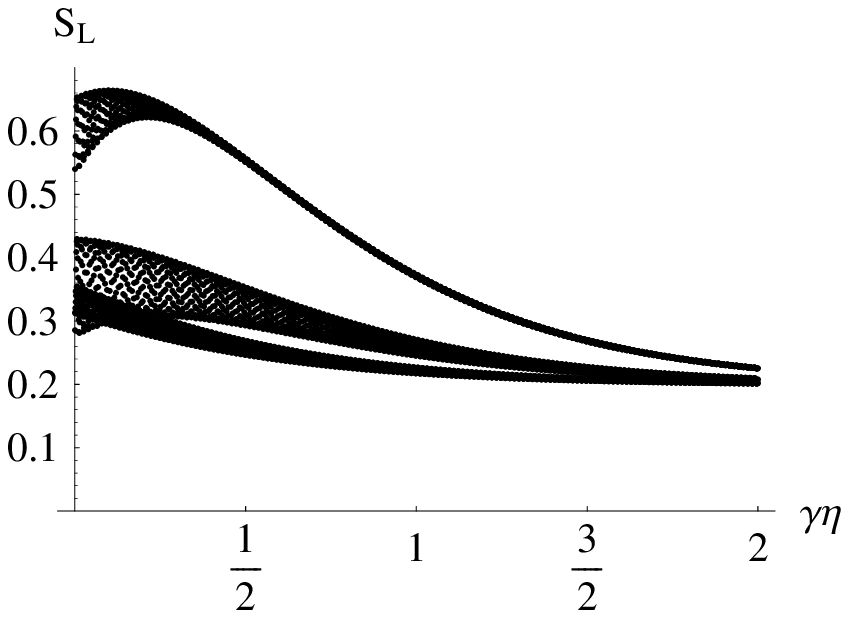}
\includegraphics[width=5cm]{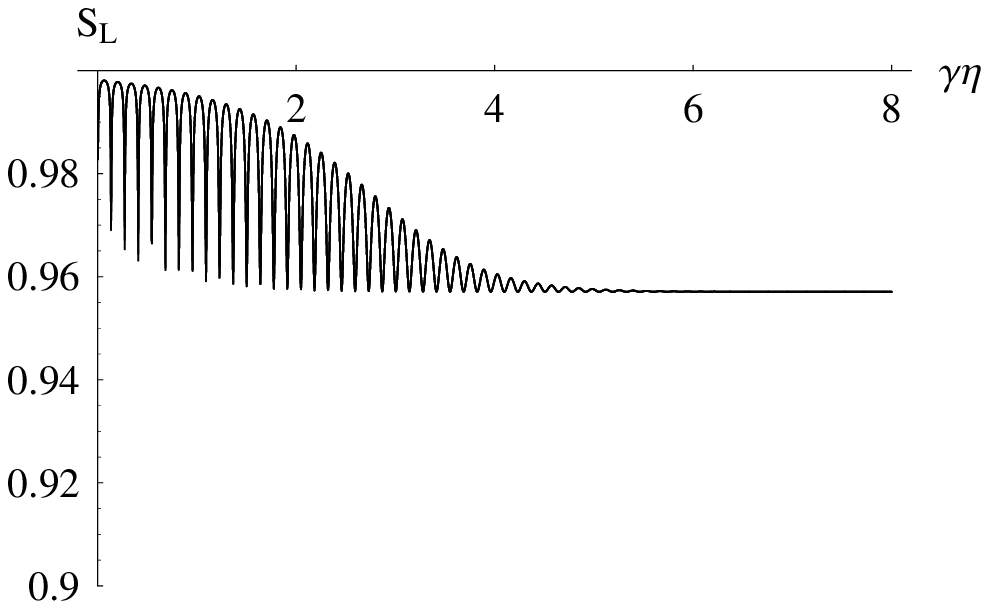}\\
\includegraphics[width=5cm]{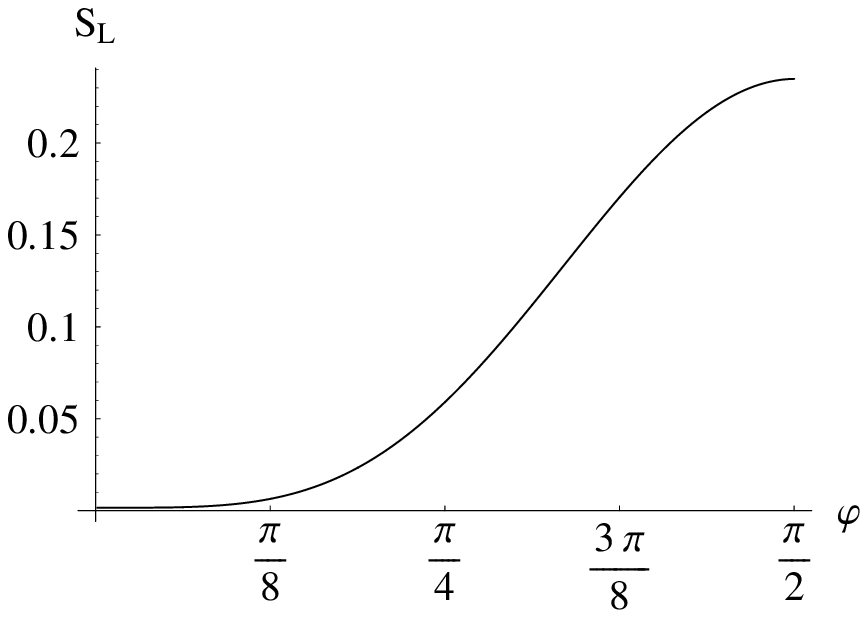}
\includegraphics[width=5cm]{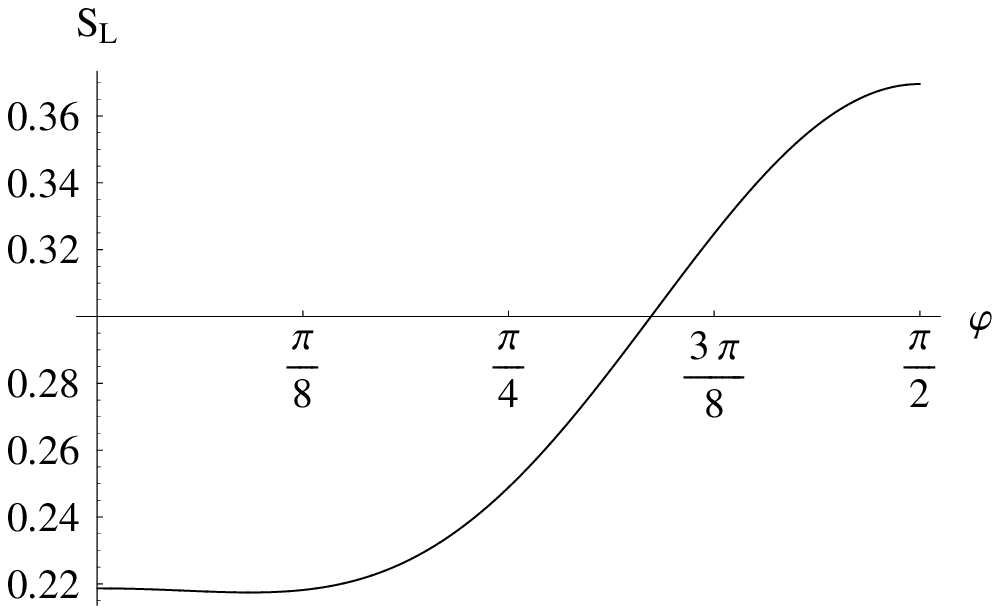}
\includegraphics[width=5cm]{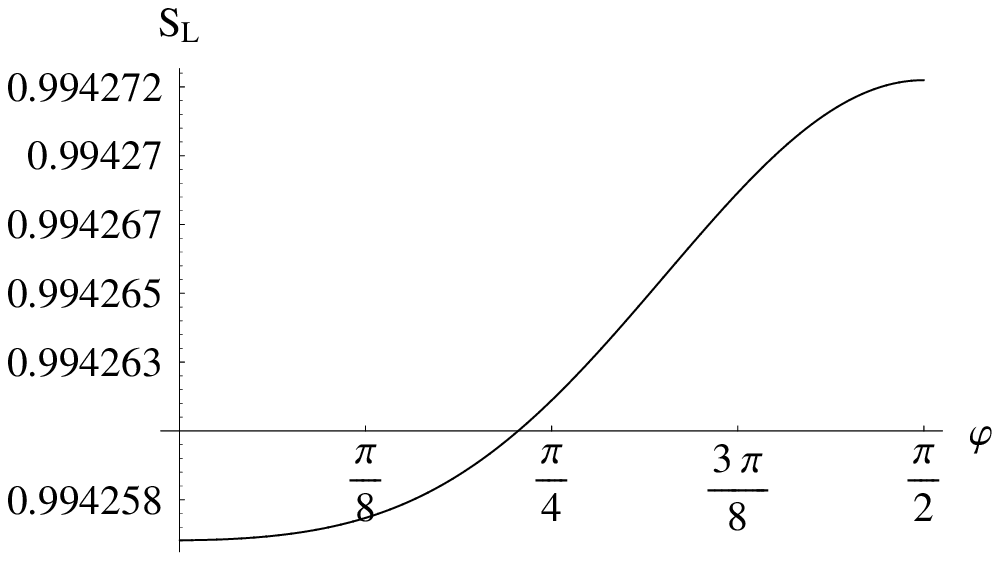}
\caption{Linear entropies $S_L =1-{\cal P}$ of $(\ref{purecat})$ with
$\Omega =2.3$, $a=2$, $m_0=\hbar=1$, $\Lambda_1=\Lambda_0=10000$,
$\delta=0$. From left to right, $\gamma = 10^{-7},10^{-4},0.1$,
respectively.
(Upper row): Evolution of $S_L$ in detector's proper time $\eta\equiv
\tau-\tau_0$ with the initial moment $\tau_0$. The three curves from top
to bottom in each plot (the bottom one in the left plot is very close to 
the $\gamma\eta$ axis, while the three curves are indistinguishable in 
the right plot) have $\varphi=\pi/2, \pi/4, 0$, respectively.
(Lower row): $S_L$ against the mixing angle $\varphi \in (0,\pi/2)$ at
the moment $\eta=1/\gamma$.}
\label{SL2}
\end{figure}

In Sec. II we stated that for a bipartite system in a pure state the
purity function measures quantum coherence in each subsystem and the
entanglement between two subsystems: a lower purity means lesser
quantum coherence in the detector or the field and stronger
entanglement between the detector and the field. To show this one may
define the linear entropy in terms of the purity as
\begin{equation}
  S_L \equiv 1-{\cal P} . \label{LiEtp}
\end{equation}
Now the value of $S_L$ is zero for a detector in a pure state and
is positive for a detector in a mixed state. By definition the linear
entropy seen by the detector will be equal to the linear entropy
seen by the field. Also, from Sec. \ref{puritypure} it is easy to
see that the greater the von Neumann entropy of the detector,
the greater $S_L$.
More precisely, for a pure state of a bipartite system, $S_L$ is a       
Schur concave function of the Schmidt vector $M^2_j$ in                    %
$(\ref{schmidt})$, and $S_L$ is invariant under unitary operations 
and decrease, on average, under local operations and classical 
communications. Thus $S_L$ is a entanglement monotone \cite{Vi00, BZ06}, 
which can serve as a measure of entanglement between the detector          %
and the field, just as good as the von Neumann entropy.                  

The linear entropies for different cases are illustrated in FIG.
\ref{SL2}. We find that, during the evolution, the UD detector with
the first excited state $\left|\right. E_1\left.\right>$ as the
initial state $(\varphi=\pi/2)$ always has a larger linear entropy,
hence a stronger entanglement with the field,  than those detectors
initially prepared in cat states or the ground state $(0 \le \varphi 
< \pi/2)$ of the free detector. 

\subsection{recoherence in ultraweak coupling regime}

In the ultraweak coupling regime ($\gamma \Lambda_1 \ll a,\Omega$)
\cite{LH2006},
\begin{eqnarray}
  \left<\right. Q^2 \left.\right>_{\rm v} &\approx&
    {\hbar\over 2\Omega m_0} \coth{\pi\Omega\over a}
    \left(1-e^{-2\gamma\eta}\right),\nonumber\\
  \left<\right. Q^2 \left.\right>_{\rm a_0} &\approx&
    {\hbar\over 2\Omega m_0} e^{-2\gamma\eta},
\label{QQweak}
\end{eqnarray}
$\left<\right. P^2 \left.\right>_{\rm v,a_0}\approx m_0^2 \Omega^2
\left<\right. Q^2 \left.\right>_{\rm v,a_0}$ and $\left<\right. P,Q
\left.\right>_{\rm v,a_0}$ are negligible, thus
\begin{equation}
  {\cal U}\approx {\hbar\over 2}\left[ e^{-2\gamma\eta}+
  \left(1-e^{-2\gamma\eta}\right)\coth {\pi\Omega\over a}\right],
\end{equation}
and 
\begin{eqnarray}
   {\cal P} &\approx&
   \left({\hbar\over 2{\cal U}}\right)^3
   e^{-4\gamma\eta}\left\{ \left[ \cos^2\varphi
   + \left(e^{2\gamma\eta}-1\right)\coth{\pi\Omega\over a}\right]^2
   \right. \nonumber\\ & & \left. +
   2\cos^2\varphi\sin^2\varphi\left[ 1 + \left(e^{2\gamma\eta}-
   1\right)\coth {\pi\Omega\over a}\right]+ \sin^4\varphi\right\}.
\label{PurWeak}
\end{eqnarray}
For $a/\Omega$ sufficiently small, one has
\begin{equation}
   {\cal P} \approx 1+2
   e^{-2\gamma\eta}\left(e^{-2\gamma\eta}-1\right)\sin^4\varphi .
\label{SE1to0}
\end{equation}
One can see that in this regime ${\cal P}$
is very close to unity at late times, when each subsystem re-gains
almost all quantum coherence and turns into a nearly pure state.

Indeed, observing the upper-left plot of FIG. \ref{SL2}, the linear
entropy $S_L$ in the detector
increases from zero right after the coupling is switched on, reaches
a maximum $(1/2)\sin^4\varphi$ at $\eta \approx \ln 2/2\gamma$, then
decays to a small common value that detectors with all other initial
states will asymptopte to. This decay of the degree of entanglement
or the restoration of the degree of quantum coherence is known as
``recoherence" \cite{AngRecoh}. From $(\ref{lateP})$ one sees that
the late-time recoherence manifests only in the ultraweak coupling
regime with sufficiently low acceleration (temperature), where the
late-time RDM of the detector looks very close to the density matrix
of the ground state of a free detector. Thus the recoherence here
characterizes the process of relaxation or spontaneous emission by 
which the detector initially in excited state will finally fall into 
a steady state which is very close to the ground state of the free 
detector. Nevertheless, {\it full} recoherence is impossible once the 
coupling is on, since the late-time linear entropy
\begin{eqnarray}
  S_L|_{\gamma\eta\to \infty} &\approx& 1-\tanh{\pi\Omega\over a} +
  \gamma {2\tanh^2{\pi\Omega\over a}\over \pi \Omega}\times\nonumber\\
  & & \left[\Lambda_1 + \ln{\Omega\over a} - {\rm Re}\left[ 
  \psi\left({i\Omega\over a}\right) +{i\Omega\over a}\psi^{(1)}\left(
  {i\Omega\over a}\right)\right]\right] + O(\gamma^2)\nonumber\\
  &\stackrel{a\ll \Omega}{\longrightarrow}& {2\gamma\over \pi\Omega}
  (\Lambda_1-1)+O(\gamma^2),
\end{eqnarray}
($\psi^{(1)}(x)\equiv d\psi(x)/dx$) remains nonzero for any positive
$\gamma$, even when $a\to 0$.

Eq. $(\ref{QQweak})$ yields
\begin{equation}
  \left<\right. Q^2 \left.\right>_{11} \approx
    {\hbar\over 2\Omega m_0}\left(1+2e^{-2\gamma\eta}\right),
\end{equation}
and $\left<\right. P^2 \left.\right>_{11} \approx m_0^2\Omega^2
\left<\right. Q^2 \left.\right>_{11}$ in the ultraweak coupling
limit. When $\varphi=\pi/2$ and $\eta=\ln 2/2\gamma$, $\left<\right.
Q^2\left.\right>_{11}|_{\eta=\ln 2/2\gamma} =(\left<\right.
Q^2(0)\left.\right>_{00}+\left<\right. Q^2(0) \left.\right>_{11})/2$
and $\left<\right. P^2\left.\right>_{11} |_{\eta=\ln
2/2\gamma}=(\left<\right. P^2(0)\left.\right>_{00}+ \left<\right.
P^2(0)\left.\right>_{11})/2$. It may seem that the intermediate state
of the detector during the transition is a cat state which is a
superposition of the ground state and the first excited state of the free
detector, but this is not true. Recall that the Einstein A coefficient
describing the spontaneous emission of an atom in a thermal bath of
photons pertains to transition probability rather than amplitude.
Large $S_L$ in transient indicates that the intermediate state during
the spontaneous emission is a mixed state, in contrast to the zero
$S_L$ for a pure cat state (at the initial moment of the middle curve
in the upper-left plot of FIG. \ref{SL2}). The value of
$\left<\right. ..\left.\right>_{11}$ in transient is mainly an
ensemble (probabilistic) average of the population in the ground
state and the population in the first excited state. Indeed,
transformed to the representation of energy eigenstates of the free
detector, the RDM of the detector becomes
\begin{eqnarray}
  \rho^R_{0,0}&\approx &\cos^2\varphi+(1-e^{-2\gamma\eta})
       \sin^2\varphi,\\
  \rho^R_{1,1}&\approx& e^{-2\gamma\eta}\sin^2\varphi, \\
  \rho^R_{0,1}&=&(\rho^R_{1,0})^* \approx
    e^{-(\gamma-i\Omega)\eta+i\delta}\cos\varphi\sin\varphi ,
\end{eqnarray}
with other elements negligible. When $\varphi=\pi/2$, it behaves like 
the population of the first excited state $\rho^R_{1,1}$ decaying to
the ground state directly while $\rho^R_{0,1}$ and $\rho^R_{1,0}$ are
always negligible.

\subsection{Beyond the ultraweak coupling regime}
\label{nMark}

As the coupling increases, $\Lambda_0$ and $\Lambda_1$ terms
\cite{LH2005, LH2006} take over (see upper-middle plot of FIG.
\ref{SL2}) and dominate (upper-right, FIG. \ref{SL2}). The emerging
initial oscillations of $S_L$ is coming from the $\Lambda_0$ term,
which corresponds to the time-scale of the switching and would be
tamed by turning on the coupling smoothly, while the large late-time
$S_L$ (or the small ${\cal P}$) is due to the $\Lambda_1$ term,
which corresponds to the time-resolution of this theory.

When  $\gamma \Lambda_1$ is comparable with or greater than $a$ and 
$\Omega$, the system is in the
non-Markovian regime and the purity is always small, which implies
that the detector experiences strong decoherence associated with
strong entanglement between the detector and the field. The behavior
of the detector is dominated by the physical cut-offs ($\Lambda_0$ and
$\Lambda_1$) and the differences between initial states of the
detector can be negligible (see upper-right plot of FIG. \ref{SL2}.)
For example, when $\varphi=\pi/2$, just like the case with the
detector initially in its ground state, the initial distribution of
the RDM in energy-eigenstate representation $\rho^R_{m,n}$ peaked at
the element $\rho^R_{1,1}$ would, upon the switch-on of the coupling,
collapse rapidly into a distribution widely spread over the whole
density matrix, for which the energy eigenstates of the free detector
cannot form a good basis because the off-diagonal terms of the RDM
do not vanish even in steady state \cite{LH2006}.
The late-time linear entropy
\begin{equation}
  S_L|_{\gamma\eta\gg 1} \approx 1- {\pi/2  \over
  \sqrt{ {2\over\Omega}\gamma\Lambda_1 {\rm Re}
  \left[ {ia \over \gamma+ i\Omega} -2i\psi_{\gamma+i\Omega}
  \right]}} + O(\Lambda_1^{-3/2})
\end{equation}
is still very close to unity. Hence there is no late-time recoherence
in this regime, where the quantum state of the combined system is
anything but a direct product of the state of the detector and that
of the field.

\begin{figure}
\includegraphics[width=5cm]{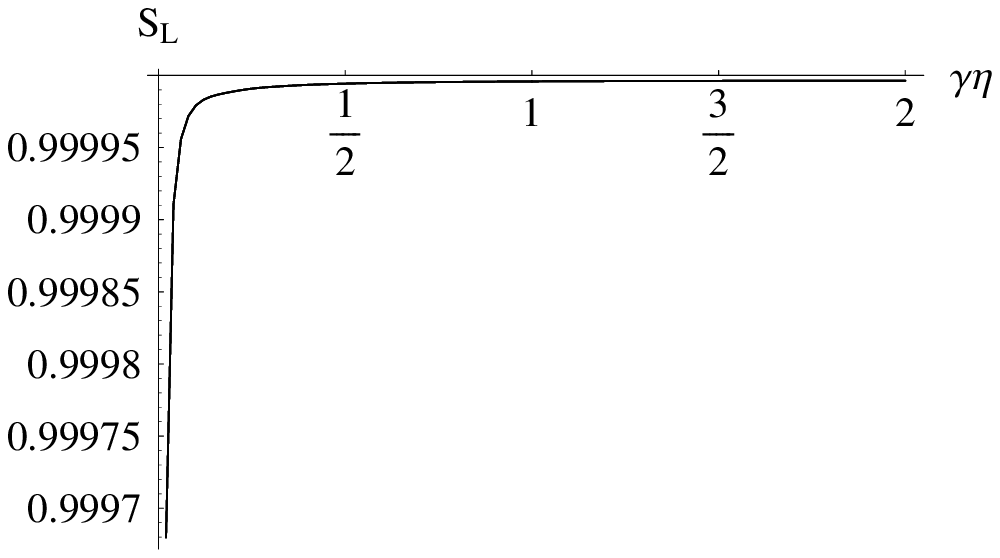}
\includegraphics[width=5cm]{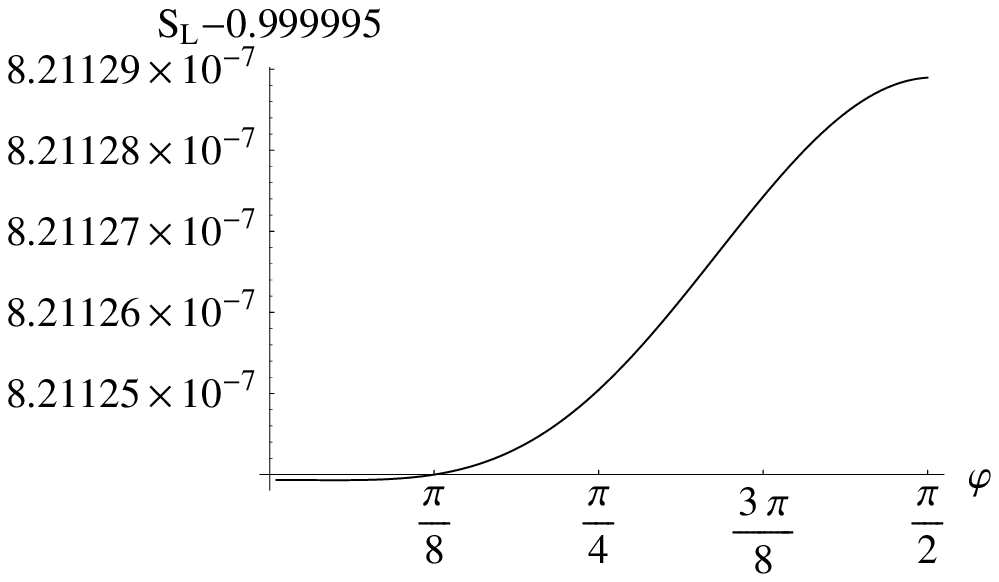}
\caption{$S_L$ in ultrahigh acceleration limit: $a=2\times 10^6$,
$\gamma =0.1$, other parameters have the same values as those in
FIG. \ref{SL2}. As indicated in the right plot, the curves with
different $\varphi$ are not distinguishable in the left and middle
plots.} \label{higha}
\end{figure}

A large linear entropy at late times also shows up at the ultrahigh
acceleration (or Unruh temperature) limit ($a\gg\gamma\Lambda_1$,
$\Omega$),
\begin{equation}
 S_L|_{\gamma\eta\gg 1}\approx 1-{\pi \Omega_r\over 2a} + O(a^{-2})
\end{equation}
(see FIG. \ref{higha}). When the coupling is weak enough, the energy 
eigenstates of the free detector can still form a good basis and the RDM 
of the detector is approximately a thermal state in energy-eigenstate
representation: all off-diagonal elements are negligible.
Note that the ultrahigh temperature limit is still in a Markovian
regime, so we see that strong entanglement does not imply a
non-Markovian process.
Although the difference between the initial and the final state
looks large in this limit, the detector would feel that
the state of the field is not heavily altered by the detector
\cite{LH2006}. This would become clear from the viewpoint of the
field which we now turn to.

\subsection{View from the field}

\begin{figure}
\includegraphics[width=5cm]{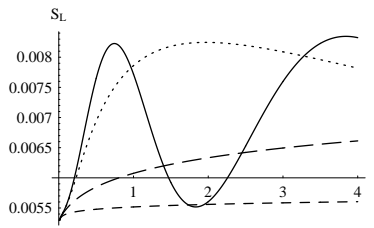}
\caption{Evolution of $S_L$ in Minkowski time $t$ after the coupling
is switched on at $t=0$. Here $\gamma=.0001$, $\Omega=2.3$,
$\Lambda_0=\Lambda_1=100$, $\varphi=\pi/4$, $a=1,4,16,64$ for the
solid, dotted, long-dashed and short-dashed curves, respectively.
One can see that the larger $a$, the
more pronounced time-dilation for $S_L$ in Minkowski time $t$.
Note that this plot is still in the weak coupling regime.}
\label{SLvst}
\end{figure}

\begin{figure}
\includegraphics[width=5cm]{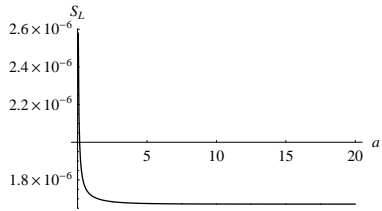}
\caption{$S_L$ against $a$ at fixed Minkowski time $t=10^6$.
The coupling is switched on at $t_0=0$.
Here $\gamma = 10^{-9}$, $\Omega=2.3$, $\Lambda_0=\Lambda_1=1000$,
$\varphi =\pi/2$, other parameters are the same as those in FIG.
\ref{SL2}.}
\label{SLvsa}
\end{figure}

Eq. $(\ref{PurWeak})$ implies that
\begin{equation}
  \left.{d S_L\over d\tau}\right|_{\tau\to \tau_0} \approx
  \gamma\left[{1\over 2}\coth{\pi\Omega\over a}
  (7-4\cos 2\varphi+\cos 4\varphi)-2\right],
\end{equation}
in the ultraweak coupling limit.
From which one sees that the larger the proper acceleration $a$ is,
the steeper is the initial rising rate for $S_L$ in the
proper time of the detector, and the larger is the linear entropy at
fixed $\tau$ in transient. In other words, for ultraweak interactions,
entanglement grows faster in transient for larger $a$ measured by
the clock of the detector. However, the detector with a larger $a$
has a more pronounced time dilation for the Minkowski observer (see
FIG. \ref{SLvst}), so the linear entropy for a larger $a$ does not
have to grow faster in Minkowski time $t$.
In fact, in FIG. \ref{SLvsa}, we see that for sufficiently small
$a$, $S_L$ at fixed Minkowski time can decrease as $a$ increases.

From the viewpoint of the field, a uniformly accelerated detector
with large $a$ has obvious changes in time only around $t=0$ because
significant time-dilation will be seen by the field once $|t|$ is
large enough.
Therefore from the viewpoint of the field most of the time the
detector actually looks frozen and the radiated power per unit
Minkowski time is small \cite{LH2005}, so the field seems not
strongly affected by the detector.


\section{Summary and Discussion}

\subsection{Summary}

Applying the non-perturbative results obtained in \cite{LH2005,
LH2006}, we analyze the evolution of quantum coherence and
entanglement in a uniformly accelerated detector - quantum field
system. By studying the evolution of a UD detector initially in a
superposition of energy eigenstates of the free detector and in 
contact with a massless scalar field in the Minkowski vacuum, 
we demonstrate that only in the
very restrictive regime of ultraweak coupling, while the detector
loses quantum coherence during spontaneous emission when the detector
makes transit from its initial state to the final state very close to
its free ground state, an {\it imperfect} recoherence will occur
eventually. This can be considered as qualitatively consistent with
the claims of \cite{AngRecoh}. In the much broader range where the
coupling is not ultraweak, our explicit calculation shows the
detector has a very low purity at late times, thus quantum coherence
never returns to the detector and quantum entanglement between the
detector and the field is always very strong.  We now discuss a few
questions and issues brought up in this investigation.

\subsection{Is there transient recoherence during spontaneous
emission?}

We have done a similar calculation for the detector initially in the
second excited state. In the ultraweak coupling limit with sufficiently
small $a/\Omega$, the evolution of  purity can be approximated by
\begin{equation}
  {\cal P} \approx 1 - 4 e^{-2\gamma \eta}(1-e^{-2\gamma \eta})  
  +6 \left[e^{-2\gamma \eta}(1-e^{-2\gamma \eta})\right]^2,
\label{SE2to0}
\end{equation}
which behaves quite similar to the case when the detector is
initially in the first excited state ($\varphi=\pi/2$ in
$(\ref{SE1to0})$ and the top curve in the upper-left plot of FIG.1.)
There exists only one extremum for $(\ref{SE2to0})$: Again when $\eta
\approx \ln 2/2 \gamma$, ${\cal P}$ has the minimal value $3/8$. Thus
there is no ``transient recoherence" or intermediate pure state
during the spontaneous emission process for the UD detector-field
system. While $\left<\right. Q^2\left.\right>_{22}|_{\eta = \ln
2/2\gamma}$ has the value of $\left<\right. Q^2(0)\left.\right>
_{11}$, which makes the detector look like it is in its first excited
state, the detector is actually in a mixture of the second, the first
and the ground state since $S_L$ is large then.

\subsection{Superposition and coherence: quantum vs classical}

The oscillating mean field emitted by the detector is a
manifestation of quantum superposition of energy eigenstates inside
the detector, but the converse is not true. An example is that if
the initial state of the detector is a superposition of its ground
state and the second excited state, the mean field emitted by the
detector is always zero.

One may ask whether the mean field carries ``quantum information"
rather than a classical one.  The expectation values of $Q$ and $P$
and the mean field actually enter into the energy conservation law
$(\ref{Econserv})$ between the internal energy of the detector and
the radiated energy of a monopole radiation emitted by the detector.
As far as the energy balance is concerned there is no essential
difference between the mean field and the classical field. From this
view the mean field carries no quantum information.

To determine the initial state of the detector from the evolving
quantum field, an observer without any {\it a priori} knowledge about
the set-up has to find out the correlation functions to all orders,
even though in the problem at hand only the one-point (mean field)
and two-point functions of the detector and those of the field are
sufficient as they contain the full information about the combined
system. This is because the UD detector theory is linear and so all
higher-order correlation functions can be written as combinations of
these lower-order functions.

\subsection{Information in a mixed state  vs. 
No information in a thermal state}

As long as the coupling between the detector and the field is on, the
detector and the field are separately in a mixed state of its own,
while the combined system remains in a pure state. The mixed state of
the detector carries information of the initial state until $\left<
Q^2\right>_a$, $\left< P^2\right>_a$ and $\left< P,Q\right>_a$ all
decay away and the detector reaches a steady state. At late times,
while the energy eigenstates of the free detector cannot form a good
basis, the RDM of the detector can be diagonalized to a Boltzmann
distribution from which one can read off the same effective
temperature as those reported in \cite{LH2006}. In this sense the
final mixed state of the detector is a thermal state containing no
initial information. This is true for both inertial and uniformly
accelerating detectors, the former case might be a surprise.

\subsection{Detector near a black hole event horizon}

The case of a uniformly accelerated detector in Minkowski space is
analogous to the case of a detector sitting outside a Schwarzschild
black hole at a fixed radial distance. In \cite{LH2005} we have shown
that the information about the initial state of the detector will be
carried in (quantum) radiation emitted and propagated to the future
null infinity  ${\cal I}_+$ of Minkowski space, and all the radiation
containing those information eventually cross the event horizon of
the detector. In the black hole case, by contrast, a quick estimate
indicates that not all, but about one half, of the radiation emitted
from the detector will enter the event horizon of the black hole,
while the remaining radiation goes outwards to ${\cal I}_+$. The
exact ratio between the inward and outward radiations depends on the
gray body factor of the black hole and could be obtained only by
detailed calculations.

Now consider the situation that, besides this fixed detector, there
is another detector carrying some information and falling into the
black hole. One may ask how information might flow from the
free-falling detector to the fixed one, and how the correlations
between these two detectors evolve. It is also interesting to look
into the way how the quantum field aids in the propagation of
information and the evolution of those correlations. We will answer
these questions in a forthcoming paper.

\subsection{Black hole information and end state}

Another avenue to connect with the black hole problem is to treat the
detector itself as an analog of the black hole. This follows
Bekenstein's observation that the black hole behaves like an ensemble
of quantum mechanical atoms \cite{Bek75, BM76, Bek97, BG02}, whose
spontaneous emissions correspond to Hawking radiation, and the energy
level of the atom is analogous to the area level of the black hole.
When they are fed with field quanta, black holes tend to absorb more
than emit energy. Indeed, Bekenstein and Meisels showed that for
black holes the Einstein B coefficient for stimulated absorption is
greater than the coefficient for stimulated emission \cite{BM76}.

Let us assume that the theory for black hole atoms does not violate
unitarity and there is no information loss, just like our UD detector
theory. Then our results for $a=0$ indicate that, if the field is
initially in a vacuum state, the information about the black hole
will be encoded in its spontaneous emission, namely, its emitted
radiation which is not exactly thermal \footnote{Note that as
different from atoms, for a real black hole its event horizon will
have an effect on such emission.}. All initial information in the
black hole will eventually go to the field at late times, while the
final state of the black hole is sustained by the vacuum fluctuations  
of the quantum field. This is consistent with the ``no-hiding
theorem" of Braunstein and Pati (with their ancilla as our quantum
field) \cite{BraPat}: no information is hidden in the correlations
between the field and the black hole. One can further verify this
statement by noting that the correlation functions $\left<\right. 
Q(\tau)\Phi(t(\tau),{\bf x})\left.\right>_{\rm a}$ 
for all ${\bf x}$ vanish at late times.

When the black hole is radiating, the black hole itself and the field
outside the black hole are each in a mixed state. Only in the
ultraweak coupling limit can they restore most of their purity at
late times. Otherwise, in the more prevalent non-Markovian regime,
the area eigenstates of the black hole cannot form a good basis, and
the entanglement between the black hole and the field is always
large. Nevertheless, in this scenario the quantum state of the
combined system is always pure due to the unitarity we assumed.

In our model the difference between the degrees of freedom of the
detector and those of the field  is put in by hand and never
disappears, and 
we cannot address whether the ground state corresponds
to a black hole remnant, an ordinary localized  mass, or nothing.
However, in Bekenstein's atom analog highly excited eigenstates
correspond to large-area black holes. Our results indicate that in
the rather general non-Markovian regime, the combined system would
evolve to a highly entangled state between the black hole and the
field, and the final state of the black hole would be a complicated
mixed state distributed widely from the ground state to the highly
excited area-eigenstates. So at late times in the broad ranged
non-Markovian regime the black hole could end up as a 
remnant with a large mean area 
with all its initial information already leaked out and
dispersed into the quantum field.\\

\noindent{\bf Acknowledgement} 
SYL wishes to thank Zheng-Yao Su
for helpful discussions. BLH appreciates the hospitality of 
Stephen Adler while visiting the Institute for Advanced Study,
Princeton in Spring '07. This work is supported in part by an NSF
Grant PHY-0601550.

\appendix

\section{RDM of the quantum field initially in the Minkowski vacuum}

The RDM of the field for the initial condition
$(\ref{initcat})$ is similar in form as $(\ref{RDMD})$:
\begin{eqnarray}
  & &\rho^R (\Phi,\Phi';t) = {e^{-\gamma_{ij}^{\rm xy}(t)\Phi^i_{\rm x}
  \Phi^j_{\rm y}} \over \sqrt{\det \left(2\pi D_{\rm xy}\right)}}
  \left\{\cos^2\varphi +\sin^2\varphi\left( \chi(t) +\alpha_{ij}^{\rm xy}(t)
  \Phi^i_{\rm x}\Phi^j_{\rm y}\right) + \right.\nonumber\\& & \,\,\,
  \left. \sin\varphi\cos\varphi\left[ \left(e^{i\delta}\beta_1^{\rm x}+
  e^{-i\delta}\beta_2^{{\rm x}*}\right)\Phi_{\rm x}+ \left(e^{-i\delta}
  \beta_1^{{\rm x}*}+ e^{i\delta}\beta_2^{\rm x}\right)\Phi'_{\rm x}
  \right]\right\}, \label{RDMF}
\end{eqnarray}
where $\Phi^j_{\rm x} = (\Phi_{\rm x}, \Phi'_{\rm x})$, $j=1,2$ and
\begin{equation}
  D_{\rm xy}(t) \equiv \left<\right.\Phi_{\rm x}(t),
    \Phi_{\rm y}(t)\left.\right>_{00}
\end{equation}
is symmetric by definition: $ \left<\right. \hat{A},\hat{B}\left.\right>
\equiv \left<\right.(\hat{A} \hat{B} + \hat{B} \hat{A})\left.\right>/2$.
Writing the inverse matrix of $D_{\rm xy}$ as $D^{\rm xy}$ so that $D_{\rm
xy}D^{\rm y z} = \delta_{\rm x}{}^{\rm z} = \delta^3({\rm x}-{\rm
z})$, the coefficients in $(\ref{RDMF})$ obey the following relations
\begin{eqnarray}
  &&\left(\gamma_{11}+\gamma_{22}+\gamma_{12}+\gamma_{21}\right)^{\rm xy} 
    ={1\over 2}D^{\rm xy},\label{gamma1}\\
  &&\left(\gamma_{11}-\gamma_{22}+\gamma_{12}-\gamma_{21}\right)^{\rm xy} 
    = -{i\over\hbar}\left<\right.\Pi^{\rm x},
    \Phi_{\rm z}\left.\right>_{00}D^{zy},\\
  && \left(\gamma_{11}+\gamma_{22}-\gamma_{12}-\gamma_{21}\right)^{\rm xy} 
    =\nonumber\\ & & \,\,\,{2\over\hbar^2}
    \left[\left<\right.\Pi^{\rm x},\Pi^{\rm y}\left.\right>_{00}
    -\left<\right.\Pi^{\rm x},\Phi_{\rm z}\left.\right>_{00}D^{\rm z z'}
    \left<\right.\Pi^{\rm y},\Phi_{\rm z'}\left.\right>_{00}\right],
    \label{gamma3}\\
  &&\left(\alpha_{11}+\alpha_{22}+\alpha_{12}+\alpha_{21}\right)^{\rm xy} 
    =  \left<\right.\Phi_{\rm z},\Phi_{\rm z'}\left.\right>_{\rm a_0}
    D^{\rm zx}D^{\rm z'y},\label{alpha1}\\
  &&\left(\alpha_{11}-\alpha_{22}+\alpha_{12}-\alpha_{21}\right)^{\rm xy} 
    = \nonumber\\ && \,\,\, {2i\over\hbar}\left[
   \left<\right.\Pi^{\rm x},\Phi_{\rm z}\left.\right>_{\rm a_0}
    D^{\rm zy}- \left<\right.\Pi^{\rm x},\Phi_{\rm z}\left.\right>_{00}
    D^{\rm zv} \left<\right.\Phi_{\rm v},\Phi_{\rm w}\left.\right>_{\rm a_0}
    D^{\rm wy} \right],\\
  &&\left(\alpha_{11}+\alpha_{22}-\alpha_{12}-\alpha_{21}\right)^{\rm xy} 
  =\nonumber\\ && \,\,\, -{4\over \hbar^2}\left[
    \left<\right.\Pi^{\rm x},\Pi^{\rm y}\left.\right>_{\rm a_0} +
    \left<\right.\Pi^{\rm x},\Phi_{\rm w}\left.\right>_{00}
    D^{\rm wz} \left<\right.\Phi_{\rm z},\Phi_{\rm z'}\left.\right>_{\rm a_0}
    D^{\rm z'w'}\left<\right.\Pi^{\rm y},\Phi_{\rm w'}\left.\right>_{00}
    \right. \nonumber\\ & & \,\,\,\left.- D^{\rm zz'}
    \left( \left<\right.\Pi^{\rm x},\Phi_{\rm z}\left.\right>_{00}
    \left<\right.\Pi^{\rm y},\Phi_{\rm z'}\left.\right>_{\rm a_0} +
    \left<\right.\Pi^{\rm x},\Phi_{\rm z}\left.\right>_{\rm a_0}
    \left<\right.\Pi^{\rm y},\Phi_{\rm z'}\left.\right>_{00}
    \right)\right],\label{alpha3}\\
  &&\left(\beta_1 + \beta_2\right)^{\rm x} =
    \left<\right.\Phi_{\rm z}\left.\right>_{10}D^{\rm z x},\\
  &&\left(\beta_1 - \beta_2\right)^{\rm x} = 
    {2i\over\hbar}\left[\left<\right.\Pi^{\rm x}\left.\right>_{10}-
    \left<\right.\Phi_{\rm z}\left.\right>_{10}
    \left<\right.\Pi^{\rm x},\Phi_{\rm z'}\left.\right>_{00}
    D^{\rm z z'}\right].
\end{eqnarray}
Here $\Pi^{\rm x}$ is the conjugate momentum of $\Phi_{\rm x}$, and
we choose
\begin{equation}
  \chi = 1-\left<\right.\Phi_{\rm z},\Phi_{\rm z'}\left.\right>_{\rm a_0}
    D^{\rm zz'}.
\end{equation}
Note that $\gamma_{11}^{\rm xy}$, $\gamma_{22}^{\rm xy}$,
$\alpha_{11}^{\rm xy}$ and $\alpha_{22}^{\rm xy}$ are symmetric under
$({\rm x} \leftrightarrow {\rm y})$, so are $\left<\right. \Phi_{\rm
x},\Phi_{\rm y}\left.\right>_{00}$, $\left<\right.\Pi^{\rm x},
\Pi^{\rm y}\left.\right>_{00}$, $\left<\right.\Phi_{\rm x},\Phi_{\rm
y} \left.\right>_{\rm a_0}$ and $\left<\right.\Pi^{\rm x},\Pi^{\rm
y}\left.\right>_{\rm a_0}$. Thus the total number of degrees of freedom
for $\gamma_{ij}^{\rm xy}$ and $\alpha_{ij}^{\rm xy}$ are exactly the
same as the number for the two-point functions of the field, as
enumerated in the above equations (note that $(\ref{gamma1})$,
$(\ref{gamma3})$, $(\ref{alpha1})$ and $(\ref{alpha3})$ are
symmetric).

\section{Purities for detector initially in the ground state}

For a UD detector initially in the ground state, the purity is
given by $(\ref{purecat})$ with $\varphi=0$:
\begin{equation}
  {\cal P} =\sqrt{\left(G^{11}+G^{22}+2G^{12}\right)^2 \over
    \det \left[G^{ij}+(G^{ij})^*\right]} =
  {\hbar/2\over {\cal U}}\label{purity00}
\end{equation}
where
\begin{equation}
  {\cal U} \equiv \sqrt{ \left<\right. P^2 \left.\right>_{00}
    \left<\right. Q^2 \left.\right>_{00} - \left(\left<\right.
    P,Q \left.\right>_{00}\right)^2 },\label{UncertFn}
\end{equation}
is the uncertainty function and  ${\cal U}\ge \hbar/2$ is the
Robertson-Schr\"odinger uncertainty relation.
The physics of the detector with such an initial state has
been studied in detail in Ref. \cite{LH2006}.

The purity of the field $(\ref{RDMF})$ for the same initial
state is
\begin{eqnarray}
 & & {\cal P}_{\Phi}|_{\varphi=0}=\left\{\det\left[{4\over \hbar^2}
    \left(\left<\right.\Phi_{\rm x}\Phi_{\rm y}\left.\right>_{00}
    \left<\right.\Pi^{\rm y}\Pi^{\rm z}\left.\right>_{00}-
    \right.\right. \right.\nonumber\\ & &\left.\left.\left. 
    \left<\right.\Phi_{\rm x}\Phi_{\rm y}\left.\right>_{00}
    \left<\right.\Pi^{\rm y},\Phi_{\rm w}\left.\right>_{00}D^{\rm ww'}
    \left<\right.\Phi_{\rm w'},\Pi^{\rm z}\left.\right>_{00}
    \right)\right]\right\}^{-1/2}.
    \label{pur00F}
\end{eqnarray}
One can compare this with $(\ref{purity00})$ to see the similarity in
appearance. Note that, while $(\ref{purity00})$ is parametrized in
the proper time of the detector $\tau$, the time variable of
$(\ref{pur00F})$ is the Minkowski time $t$, and ${\rm x}$, ${\rm y}$,
${\rm z}$, ${\rm w}$ in the above expression are defined in the whole
Minkowski-time slice rather than in the Rindler R-wedge only since
the initial state of the field is the Minkowski vacuum.

Although $(\ref{pur00F})$ looks a bit complicated for an analytic
expression, $(\ref{PPequal})$ guarantees that $(\ref{purity00})$
and $(\ref{pur00F})$ have the same value when $t= a^{-1}\sinh a\tau$.

\section{Comparison with Anglin et al. \cite{AngRecoh} }

The authors of Ref. \cite{AngRecoh} considered a model with a
harmonic oscillator at rest interacting with a scalar field in
(1+1)D. While the appearance of their action looks quite different
from $(\ref{Stot1})$, in reality their harmonic oscillator is acting
like a UD detector initially in contact with an ohmic bath. Thus when
focusing on the detector, we can borrow our results in Ref.
\cite{LH2006} for UD detector at rest ($a=0$) to describe the
behavior of their harmonic oscillator with a step switching function.

Ref. \cite{AngRecoh} assumes that the combined system has the initial
state:
\begin{equation}
 \left|\right. \psi(\tau_0)\left.\right> =
  \left( c_+ \left|\right.\alpha\left.\right> +
  c_- \left|\right.-\alpha\left.\right>\right) \otimes
  \left|\right. 0_M \left.\right>, \label{initAng}
\end{equation}
where $\left|\right.\alpha\left.\right>$ is a coherent state of the
detector, which is an eigenstate of the lowering operator $\hat{a}$:
\begin{equation}
  \hat{a}\left|\right.\alpha\left.\right> = \alpha
  \left|\right.\alpha\left.\right> .
\end{equation}
Here the eigenvalue is
\begin{equation}
  \alpha = \sqrt{m_0\Omega_r\over 2\hbar} q_0
\end{equation}
with the initial mean value $\left<\right.\alpha\left|\right.\hat{Q}
(\tau_0)\left|\right.\alpha\left.\right> =q_0$ of the detector. (Note
a transcription oversight in Ref. \cite{LH2005}: $\alpha$ should
carry the $\sqrt{m_0}$ factor.) The normalization condition is
\begin{equation}
  1=\left<\right. \psi (\tau_0) | \psi(\tau_0)\left.\right> =
|c_+|^2 + |c_-|^2 + (c_+^* c_- + c_+ c_-^*)e^{-m_0 \Omega_r q_0^2/\hbar}.
\end{equation}
If we require $|c_+|^2 + |c_-|^2=1$, then $c_+^* c_- + c_+ c_-^*$
must vanish; we will choose
\begin{equation}
  c_+ = {1\over\sqrt{2}}, \,\,\,\,\, c_- = {i\over\sqrt{2}}.
\end{equation}

Define
\begin{equation}
  \psi_{\pm}[Q, \Phi; \tau_0] \equiv \left<\right.Q,\Phi |
  \pm\alpha, 0_M \left.\right> .
\end{equation}
Since $\left|\right.\pm \alpha\left.\right>$ and $\left|\right.0_M
\left.\right>$ are Gaussian states, the RDM of the
detector must have the form
\begin{equation}
  \rho^R(Q,Q';\tau) = |c_+|^2 \rho_{++} + |c_-|^2 \rho_{--}
  + c_+^* c_- \rho_{+-} + c_+ c_-^*\rho_{-+},
\end{equation}
where
\begin{eqnarray}
  \rho_{\pm\pm}(Q,Q';\tau)&=&\int {\cal D}\Phi \psi_{\pm}[Q,\Phi;\tau]
    \psi^*_{\pm}[Q',\Phi;\tau] \nonumber\\ 
    &=& \exp\left\{ -G^{ij}(\tau) Q_i Q_j \pm
    J^i_+(\tau) Q_i + F_{\pm\pm}(\tau)\right\}, \\
 \rho_{\pm\mp}(Q,Q';\tau)&=&\int {\cal D}\Phi \psi_{\pm}[Q,\Phi;\tau]
    \psi^*_{\mp}[Q',\Phi;\tau] \nonumber\\ &=& 
    \exp\left\{ -G^{ij}(\tau) Q_i Q_j \mp
    J^i_-(\tau) Q_i + F_{\pm\mp}(\tau)\right\}. \label{Rho+-}
\end{eqnarray}
The coefficients $G^{ij}$ are given by $(\ref{G1})$-$(\ref{G3})$, and
\begin{eqnarray}
  J^1_\pm +J^2_\pm &=& 2{\left<\right.Q\left.\right>_{\pm}\over
    \left<\right.Q^2\left.\right>_{00}},\\
  J^1_\pm-J^2_\pm &=& {2i\over\hbar} \left[
    \left<\right.P\left.\right>_{\pm}
    - {\left<\right.P,Q\left.\right>_{00}\over
      \left<\right.Q^2\left.\right>_{00}}
      \left<\right.Q\left.\right>_{\pm}\right],
\end{eqnarray}
where
\begin{equation}
  \left<\right. {\cal O} \left.\right>_{\pm} \equiv
  {Tr[ \hat{\cal O}\rho_{\pm +}]\over Tr\,\rho_{\pm +}},
\end{equation}
The normalization conditions give
\begin{eqnarray}
  e^{F_{++}}=e^{F_{--}}&=& {Tr\,\rho_{++}\over \sqrt{2\pi
    \left<\right.Q^2\left.\right>_{00}}}
    \exp {-\left<\right.Q\left.\right>_+^2 \over
    8\left<\right.Q^2\left.\right>_{00}},  \\
  e^{F_{-+}}=e^{F_{+-}}&=& {Tr\,\rho_{-+}\over \sqrt{ 2\pi
    \left<\right.Q^2\left.\right>_{00}} }
    \exp {-\left<\right.Q\left.\right>_-^2 \over
    8\left<\right.Q^2\left.\right>_{00}}, \label{offdiag}
\end{eqnarray}
with $Tr\,\rho_{++}=1$ and $Tr\,\rho_{-+}=e^{-m_0\Omega_r q_0^2/\hbar}$.

\begin{figure}
\includegraphics[width=5cm]{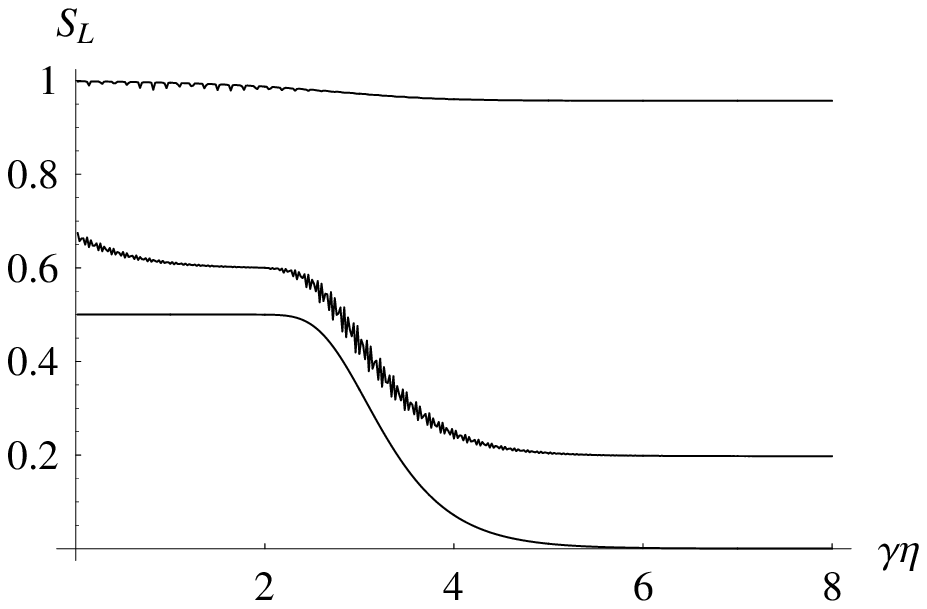}
\includegraphics[width=5cm]{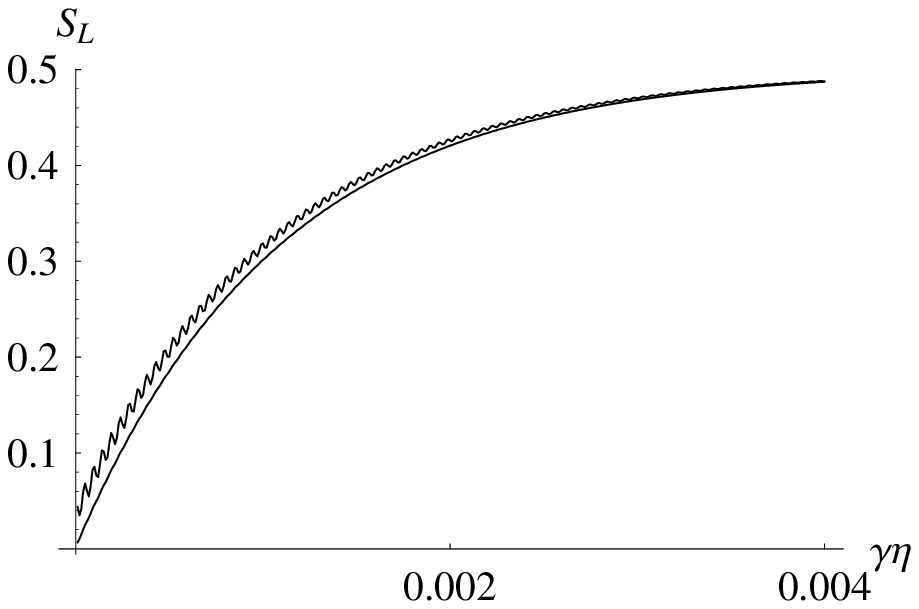}
\caption{Evolution of linear entropies $S_L=1-{\cal P}$ of
$(\ref{purCohe})$ with different parameters.
(Left) $\gamma=10^{-7}$(upper curve), $10^{-4}$(middle)
and $0.1$(lower), $q_0=10$, $\Omega =2.3$, $a=0$, $m_0=\hbar=1$,
$\Lambda_1=\Lambda_0=10000$. One can compare the lower curve with
FIG.2 in Ref. \cite{AngRecoh}.
(Right) $\gamma=10^{-7}$, $\Lambda_1=\Lambda_0=100$(lower curve) and
$2000$(upper), with other parameters the same as
before. This plot shows that the growing rate of $S_L$ right after the
coupling is switched on is almost independent of the UV cut-off in
the ultraweak coupling limit.}
\label{95a}
\end{figure}

For $c_+ = 1/\sqrt{2}$ and $c_- = i/\sqrt{2}$, the purity reads
\begin{eqnarray}
   & &{\cal P} = {\hbar\over 4{\cal U}}\left\{ 1+
    \exp\left[ -{\left< Q^2\right>_{00}\over {\cal U}^2}
    \left( \left< P \right>_+ -
    {\left< P,Q \right>_{00}\over
    \left< Q^2 \right>_{00}}
    \left< Q \right>_+\right)^2 -
    {\left< Q \right>_+^2\over
    \left< Q^2 \right>_{00}}\right]\right. +
    \nonumber\\ & & \left. \left(
    \exp\left[ -{\left< Q^2 \right>_{00}\over {\cal U}^2}
    \left( \left<P\right>_- -
    {\left< P,Q\right>_{00}\over
    \left<Q^2\right>_{00}}
    \left< Q \right>_-\right)^2 -
    {\left< Q\right>_-^2\over
    \left< Q^2\right>_{00}}\right]-1\right) 
    e^{-{2\over\hbar}m_0\Omega_r q_0^2}\right\} \nonumber\\
\label{purCohe}
\end{eqnarray}
In the Heisenberg picture, it is easy to obtain \cite{LH2005}
\begin{eqnarray}
  \left<\right.Q\left.\right>_\pm &=& {\left<\right.\alpha \left.
    \right|\hat{Q}(\tau)\left|\right. \pm \alpha\left.\right>\over
    \left<\right.\alpha \, | \pm \alpha\left.\right>} =
    {q_0\over 2}\left( q^{a*} \pm q^a \right), \\
\left<\right.P\left.\right>_\pm &=& {\left<\right.\alpha \left.
    \right|\hat{P}(\tau)\left|\right. \pm \alpha\left.\right> \over
    \left<\right.\alpha \, | \pm \alpha\left.\right>} = m_0
    {q_0\over 2}\left( \dot{q}^{a*} \pm \dot{q}^a \right),
\end{eqnarray}
where $q^a$ has been given in $(\ref{qa})$. Evolutions of linear
entropies $S_L = 1-{\cal P}$ with different parameters have been
illustrated in FIG. \ref{95a}.

In the ultraweak coupling limit, the purity $(\ref{purCohe})$ becomes
\begin{eqnarray}
  {\cal P} &\approx& {1\over 2}\left( 1+ \exp\left[-{2\over\hbar} m_0
    \Omega q_0^2 e^{-2\gamma \tau}\right] \right) -{1\over 2}
    e^{-2m_0\Omega q_0^2/\hbar}\nonumber\\ & & + {1\over 2}\exp\left[
    {2\over \hbar}m_0\Omega q_0^2\left(e^{-2\gamma \tau}-1\right)\right].
\label{finalP}
\end{eqnarray}
The second line decays from $1/2$ to $0$ in a time scale
$\tau_{\rm decoh} \sim 1/\gamma m_0 \Omega q_0^2$, while the first
line grows from $1/2$ to $1$ in a time scale $\tau_{\rm recoh}=
(2\gamma)^{-1}\ln(2m_0\Omega q_0^2/\ln 2)$.

When $m_0\Omega q_0^2$ is large, $\tau_{\rm recoh}\gg \tau_{\rm
decoh}$, so after the coupling is switched on, the purity drops from
$1$ to $1/2$ rapidly in a time scale $\tau_{\rm decoh}$, stay at
$1/2$ for a relatively long time scale $\tau_{\rm recoh}$, then goes
back to $1$ slowly. The evolution of the linear entropy in this limit
has been shown in FIG. \ref{95a} (the lower curve in the left plot),
which looks virtually the same as the ``entropy of oscillator" in
FIG. 2 of Ref. \cite{AngRecoh}.

The recoherence occurs at $\eta\approx \tau_{\rm recoh}$, when the
separation between the two wave packets become comparable to the
width of each wave packet ($q_0 e^{-\gamma\tau_{\rm recoh}}=
\sqrt{\left<\right.Q^2( \tau_{\rm recoh})\left.\right>} =
\sqrt{\hbar/m_0\Omega}$.) Before $\tau_{\rm recoh}$ the two coherent
states $\varphi_\pm(Q;\tau)$ with $\varphi_\pm(Q;\tau_0)=\left<\right.Q |
\pm\alpha \left.\right>$ are nearly orthogonal to each other and from
$(\ref{Rho+-})$ and $(\ref{offdiag})$ $\rho_{+-}$ and $\rho_{-+}$ are
suppressed so the RDM looks like $\rho^R(Q,Q')\approx (1/2)\varphi_+
(Q)\varphi_+^*(Q')+(1/2)\varphi_-(Q)\varphi_-^*(Q')$, which yields
${\cal P}\approx 1/2$.

Note that, unlike the time scales in $(\ref{SE1to0})$ and
$(\ref{SE2to0})$, here $\tau_{\rm decoh}$ and $\tau_{\rm recoh}$
depend on $m_0$ and $\Omega$ in addition to $\gamma$.

In \cite{AngRecoh} the authors reported that the entropy initially grows    
in a decoherence time corresponding to the cut-off time scale, even
though they were working in the weak coupling limit. On this point we
disagree with them since our $\tau_{\rm decoh} \sim 1/\gamma m_0
\Omega q_0^2$ is virtually independent of any cut-off in this limit
[see also FIG. \ref{95a}(right)].

Beyond the ultraweak coupling limit, the linear entropy is always
close to $1$ and never decays to $0$, so the late-time recoherence
never occurs in non-Markovian regime or the ultrahigh acceleration
regime (the upper and middle curves in FIG. \ref{95a} (left)). Again
the late-time value of the purity of the detector is $(\ref{lateP})$,
quite independent of the initial states of the detector.

\end{document}